\documentclass[a4paper]{revtex4}
\usepackage[top=1in, bottom=1in, left=1in, right=1in]{geometry}
\usepackage{amsmath}
\usepackage{amssymb}
\usepackage{amsfonts}
\usepackage{graphicx}
\usepackage{dcolumn}
\usepackage[colorlinks,linkcolor=blue,citecolor=blue,urlcolor=blue]{hyperref}

\begin{document}

\title{Preinflationary dynamics of power-law potential in loop quantum cosmology}
\author{M. Shahalam \thanks{E-mail address: shahalam@zjut.edu.cn}}
\affiliation{Institute for Advanced Physics $\&$ Mathematics, Zhejiang University of Technology, Hangzhou, 310032, China}
%%%%%%%%%
\begin{abstract}
In this article, I mainly discuss the dynamics of the pre-inflationary Universe for the potential $V(\phi) \propto \phi^n$ with $n=5/3$ in the context of loop quantum cosmology, in which the big bang singularity is resolved by a non-singular quantum bounce. In the case of the kinetic energy-dominated initial conditions of the scalar field at the bounce, the numerical evolution of the Universe can be split up into three regimes: {\em  bouncing, transition,} and {\em slow-roll inflation}. In the bouncing regime, the~numerical evolution of the scale factor does not depend on a wide range of initial values, {or on}~the inflationary potentials. I calculate the number of $e$-folds in the slow-roll regime, by which observationally identified initial conditions are obtained. Additionally, I display the phase portrait for the model under consideration.
\end{abstract}

%\keyword{Inflation; Cosmology; Loop quantum gravity}

\maketitle
%%%
\section{Introduction}
\label{intro}
Cosmic inflation is a very popular paradigm in the early Universe. It resolves various important issues in cosmology, such as the horizon and flatness  issues. Further, the theory of cosmic inflation describes the origin of inhomogeneities which are observed in the cosmic microwave background and the structure formation of the Universe~\cite{guth1981,staro1980}.
A wide range of the inflationary models are consistent with the present observations. According to Planck 2015 data, the quadratic potential  is almost ruled out in comparison with the power-law $(V(\phi) \propto \phi^n$ with $n<2)$ and Starobinsky models~\cite{Planck2015}. Therefore, in this article, I shall choose the potential $V(\phi) \propto \phi^{5/3}$.

In classical general relativity (GR), all inflationary models  {struggle}%was "suffer", please confirm.
~with the{ inevitable} big bang singularity%was " that is inevitable"
~\cite{borde1994,borde2003}. One way to address this issue  is to work in loop quantum cosmology (LQC), in which the big bang singularity is replaced by a quantum bounce~\cite{ashtekar2011,ashtekar2015,barrau2016,yang2009}. LQC gives an important  explanation of cosmic inflation and the pre-inflationary dynamics~\cite{agullo2013a,agullo2013b,agullo2015}. {In the literature, different~approaches are studied for the pre-inflationary Universe and the cosmological perturbations, namely dressed metric, deformed algebra,  hybrid quantization, etc. These approaches provide the same set of dynamical equations for background, but their perturbations are different. In the context of the hybrid approach, some uniqueness criteria are described in order to remove some ambiguities in the quantization of models~\cite{a1,a2,a3,a4}, which are consistent with observations~\cite{a3,a5}. In Reference~\cite{agullo2015}, the~post-bounce regime of the slow-roll inflation for the primordial power spectrum of tensor and scalar perturbations are discussed, whereas the effective equations of evolution for the scalar and tensor perturbations are distinct as the discrepancy seems {to be in the mass}%was "in mass", please confirm.
~term, which arises due to the different quantization procedures~\cite{a6}.} It is fascinating  to see that the Universe which starts at the quantum bounce generically enters the slow-roll inflation~\cite{ashtekar2010,psingh2006,Bonga2016,Tao2017,alam2017,alam2018}.
In this article, I am mainly concerned with the background evolution of the Universe in LQC. Moreover, when the kinetic energy of the scalar field initially  dominates at the bounce, the numerical evolution of the Universe splits up into three important regimes prior to preheating: {\em bouncing, transition,} and {\em slow-roll inflation}.

{The quadratic potential ($n=2$) has been almost disfavored by the Planck 2015 data~\cite{Planck2015}. Therefore, it increases the curiosity in studying the inflation for the above potential with $n<2$. The detailed  investigations  for the power-law potential with $n=7/4,4/3,1,2/3$, and $1/3$ have been  studied in my previous paper~\cite{alam2017}. In this conference article, I report my study for $n=5/3$, and find the observationally consistent initial conditions of the inflaton field  that provide the slow-roll inflation together with at least 60 or more $e$-folds. Although the results are similar to those presented in~\cite{alam2017}, it is not {taken} for granted, as we already know, for example, that $n=2$ is not consistent with data.} This~article is based on my previous work~\cite{alam2017}. However, the background evolution with the quadratic and Starobinsky potentials have been examined in~\cite{Tao2017}. In the present article, the physically viable initial conditions of the inflaton field, in the case of $n=5/3$, are identified that are consistent with~observations.
%%%%%%%%
\section{Background Evolution and Phase Space Analysis}
\label{sec:EOM}
In the context of LQC, the modified Friedmann equation in  a flat Friedmann--Lemaitre--Robertson--\\Walker Universe can be written as~\cite{ashtekar2006}:
\begin{eqnarray}
H^2=\frac{8 \pi}{3 m_{pl}^2}~\rho \Big{(}1-\frac{\rho}{\rho_c}\Big{)}.
\label{eq:H}
\end{eqnarray}

Here, $H=\dot{a}/a$ is the Hubble parameter, $m_{pl}$ is the Planck mass, and $\rho$ and $\rho_c$ are the inflaton and the critical energy densities, respectively. The numerical value of the critical energy density is given as $\rho_c \simeq 0.41 m_{pl}^4$~\cite{Meissne,Domagala}.

The Klein--Gordon equation in LQC does not change, and can be written as:
\begin{eqnarray}
\ddot{\phi}+3H \dot{\phi}+ \frac{dV(\phi)}{d\phi}=0.
\label{eq:ddphi}
\end{eqnarray}

From Equation~(\ref{eq:H}), one can notice that $H=0$ at $\rho=\rho_c$,  which means that the bounce occurs at $\rho=\rho_c$. The background evolution with the bouncing regime has been studied extensively. One of the main results is that beginning from the bounce, a desired slow-roll inflation is achieved~\cite{psingh2006,Tao2017,ashtekar2011,alam2017,alam2018}. Keeping this in mind, we shall discuss ``bounce and the slow-roll inflation'' with the following potential:
\begin{eqnarray}
V(\phi)=\frac{1}{2}m^{4-n}\phi^n.
\label{eq:pot}
\end{eqnarray}

Here, $m$ represents the mass dimension. I am interested in a particular value of $n$: $n = 5/3$. The~corresponding value of $m$ is $ m=10^{-5}m_{pl}$, which is consistent with the observation. The similar work in case of the power-law potential with different values of $n$ has been studied in~\cite{alam2017}.

At the quantum bounce ($t=t_B$), $\rho = \rho_c$
\begin{eqnarray}
\frac{1}{2}\dot{\phi}^2(t_B)+V(\phi(t_B))&=& \rho_c ~~\text{and}~~ \dot{a}(t_B)=0.
\label{eq:bounce} \end{eqnarray}
This implies that
\begin{eqnarray}
\dot{\phi}(t_B) &=& \pm \sqrt{2 \Big{(} \rho_c - V(\phi(t_B)) \Big{)}}, 
\label{eq:bounce2}
\end{eqnarray}
and a convenient choice for $a(t_B)$ is $a(t_B) = 1$.
Hereafter, we shall read $\phi(t_B)$, $\dot{\phi}(t_B)$, and $a(t_B)$ as $\phi_B$,  $\dot{\phi}_B$, and $a_B$. Let us first present the following parameters, which are vital in this article~\cite{alam2017}.
%%%%%%
\begin{enumerate}
\item The equation of state $w(\phi)$ for inflaton:
\begin{eqnarray}
w(\phi) = \frac{\dot{\phi}^2/2-V(\phi)}{\dot{\phi}^2/2+V(\phi)} \simeq -1, \text{during the slow-roll inflation}
\label{eq:w}
\end{eqnarray}
%%%%%%
\item The slow-roll parameter $\epsilon_H$:
\begin{eqnarray}
\epsilon_H = - \frac{\dot{H}}{H^2} \ll 1, \text{during the slow-roll inflation}
\label{eq:epsilon}
\end{eqnarray}
%%%%%%%
\item The number of $e$-folds $N_{inf}$ during the slow-roll inflation:
\begin{eqnarray}
N_{inf} = ln \Big{(} \frac{a_{end}}{a_i} \Big{)} =  \int_{t_i}^{t_{end}} H(t) dt 
 = \int_{\phi_i}^{\phi_{end}} \frac{H}{\dot{\phi}} d\phi \simeq \int_{\phi_{end}}^{\phi_i} \frac{V}{V_{\phi}} d\phi
\label{eq:Ninf}
\end{eqnarray}
%%%%%%
\item In the bouncing regime, one can find an analytical solution of the expansion factor $a(t)$~\cite{alam2017}:
\begin{eqnarray}
a(t) &=& a_B \left( 1+ \delta \frac{t^2}{t_{pl}^2} \right)^{1/6},
\label{eq:a}
\end{eqnarray}
where $t_{pl}$  is the Planck time and $\delta = \frac{24 \pi \rho_c}{m_{pl}^4}$ denotes a dimensionless constant.
\end{enumerate}

For potential (\ref{eq:pot}) with $n=5/3$, we shall use only the positive values of the scalar field $\phi$ in order to get the real potential. Let us evolve the background Equations (\ref{eq:H}), (\ref{eq:ddphi}) with (\ref{eq:pot}) and $n=5/3$. At the bounce, we shall consider only the $\dot{\phi}_B > 0$ case, as similar analysis can also be done for $\dot{\phi}_B < 0$. Further, initial values of the inflaton field can be divided into the kinetic energy-dominated (KED) and potential energy-dominated (PED) cases at the quantum bounce.

\begin{figure}

\begin{center}
\begin{tabular}{ccc}
{\includegraphics[width=0.46\textwidth]{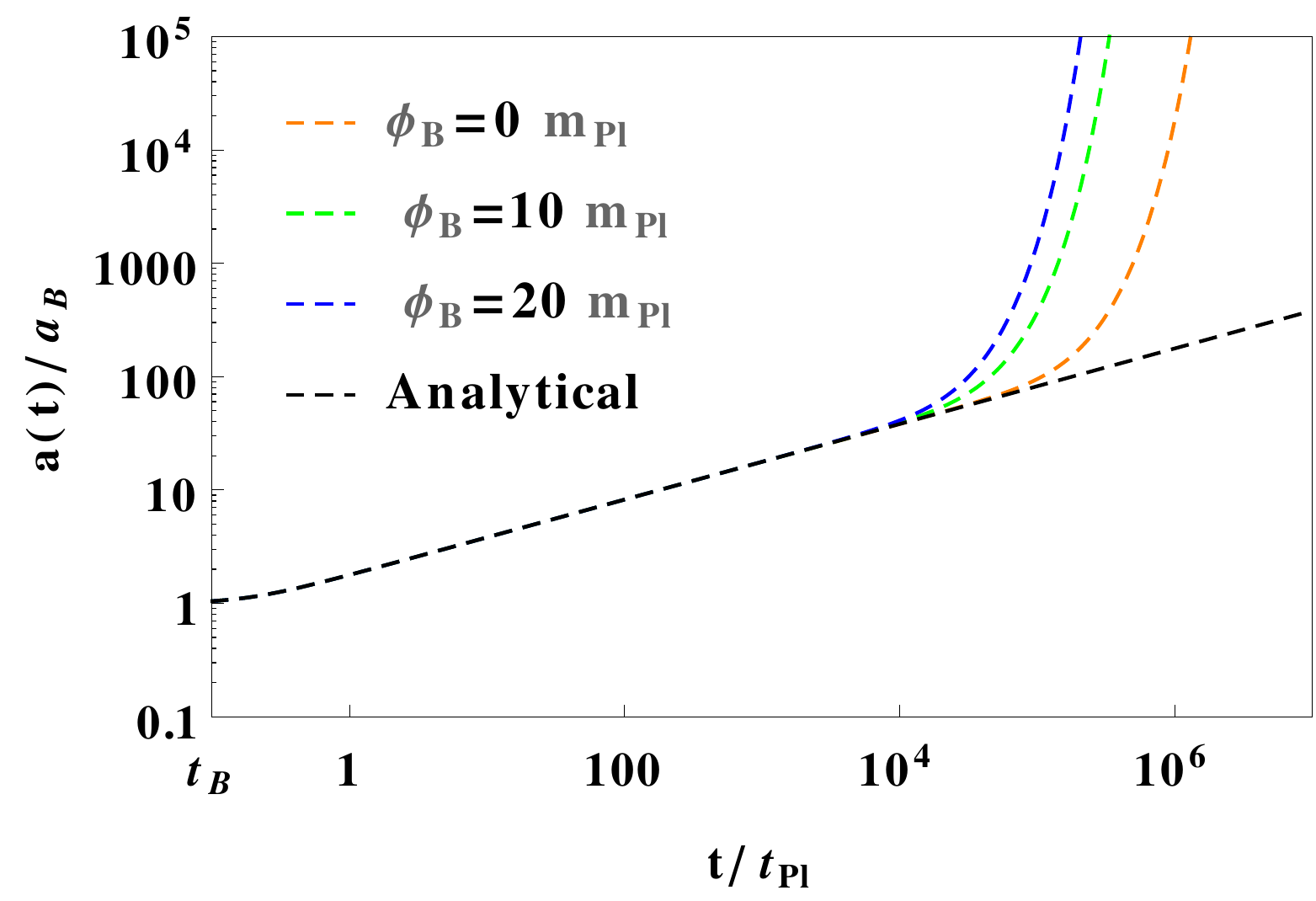}} &
{\includegraphics[width=0.46\textwidth]{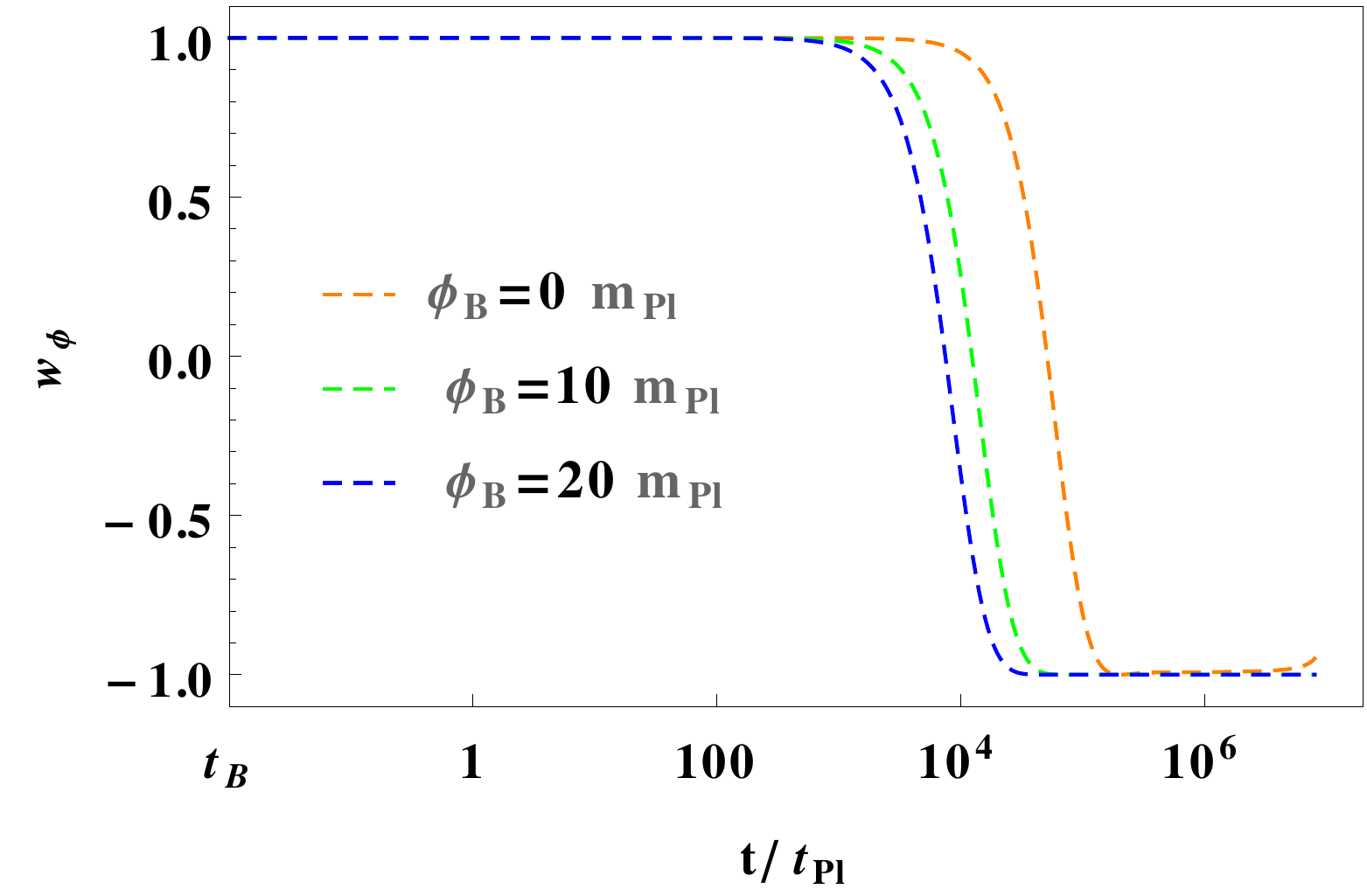}} \\\vspace{6pt}
{(\textbf{a})} & {(\textbf{b})}\\

{\includegraphics[width=0.46\textwidth]{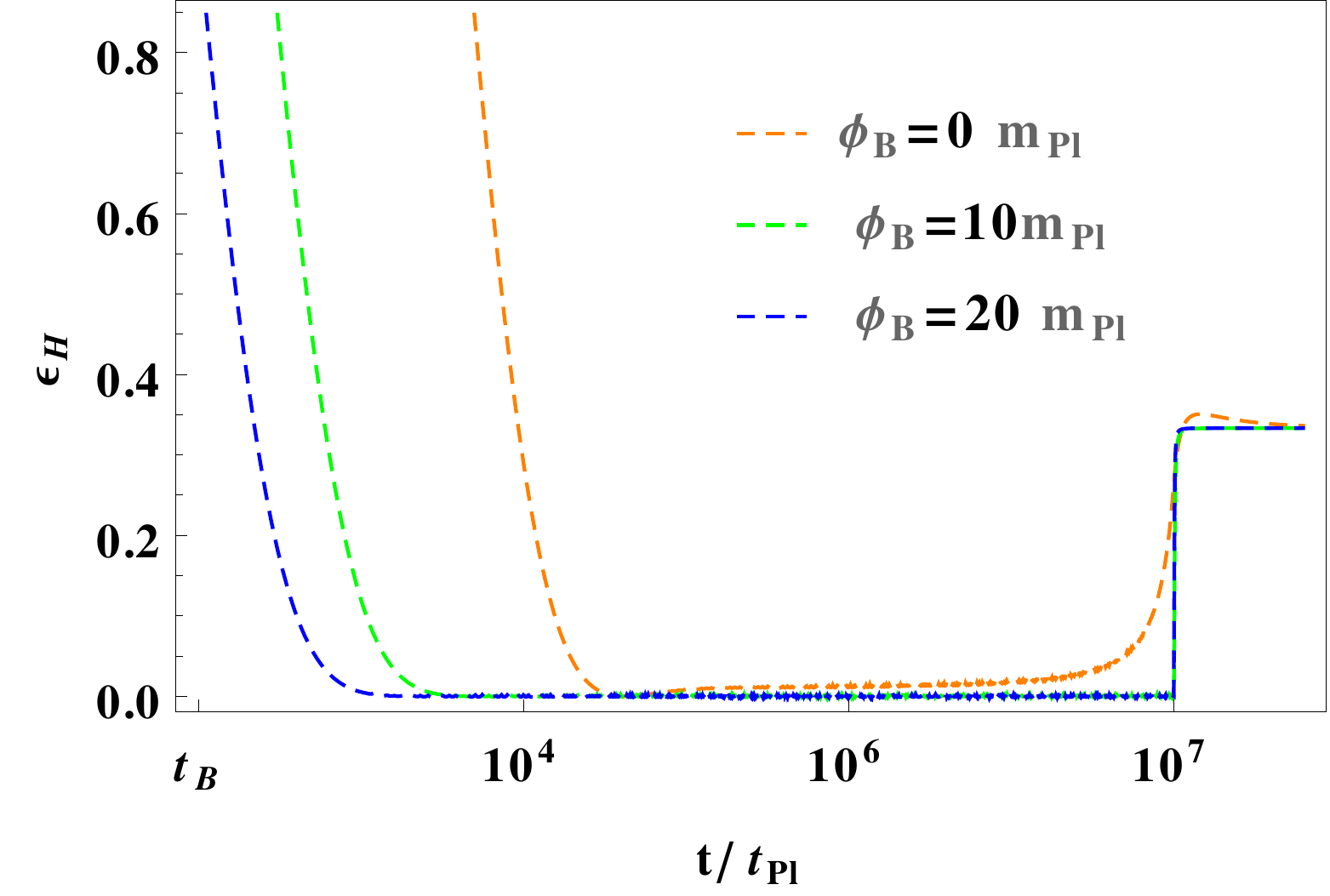}}
&
{\includegraphics[width=0.46\textwidth]{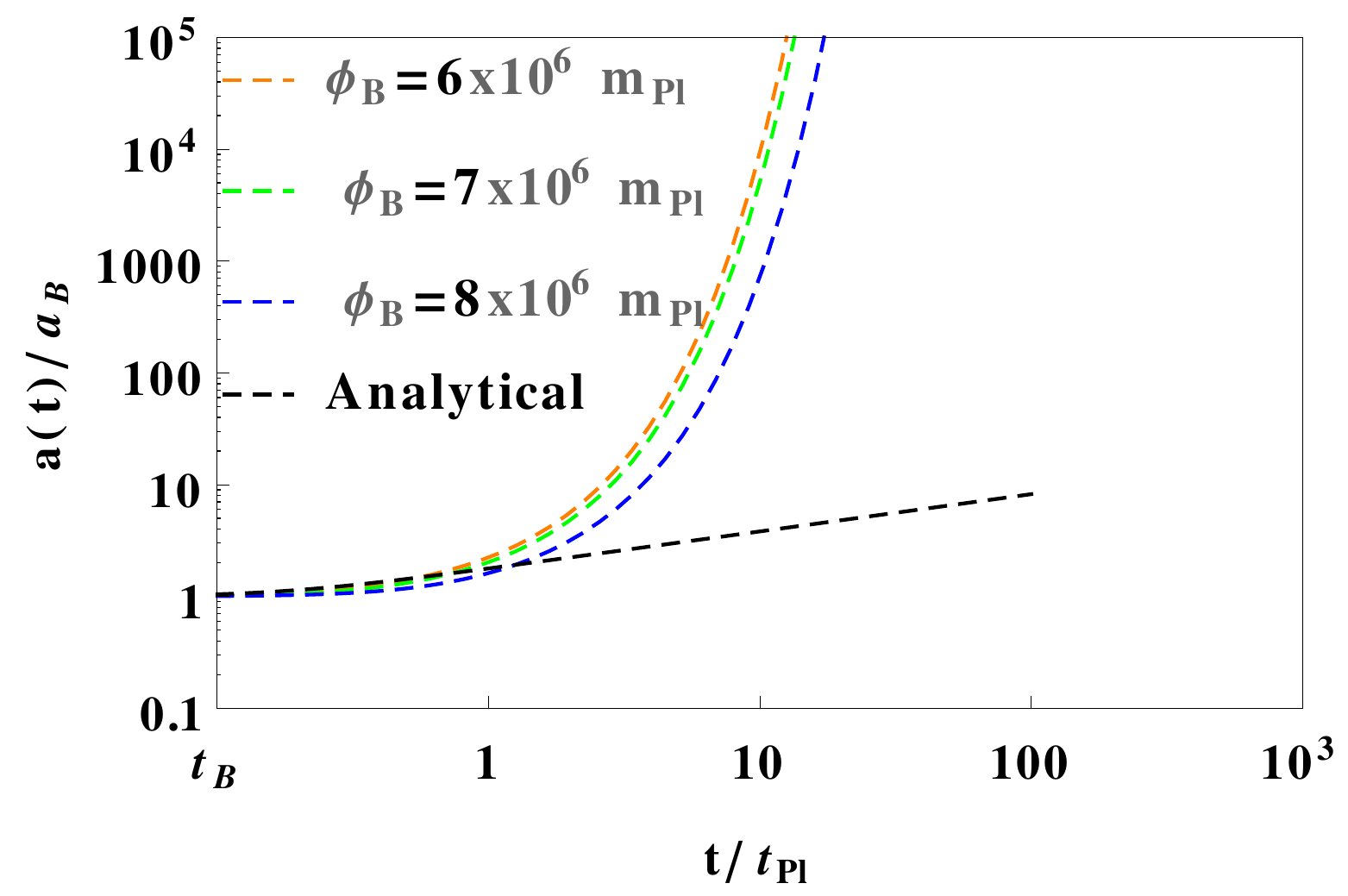}}  \\\vspace{6pt}
{(\textbf{c})} & {(\textbf{d})}\\

{\includegraphics[width=0.46\textwidth]{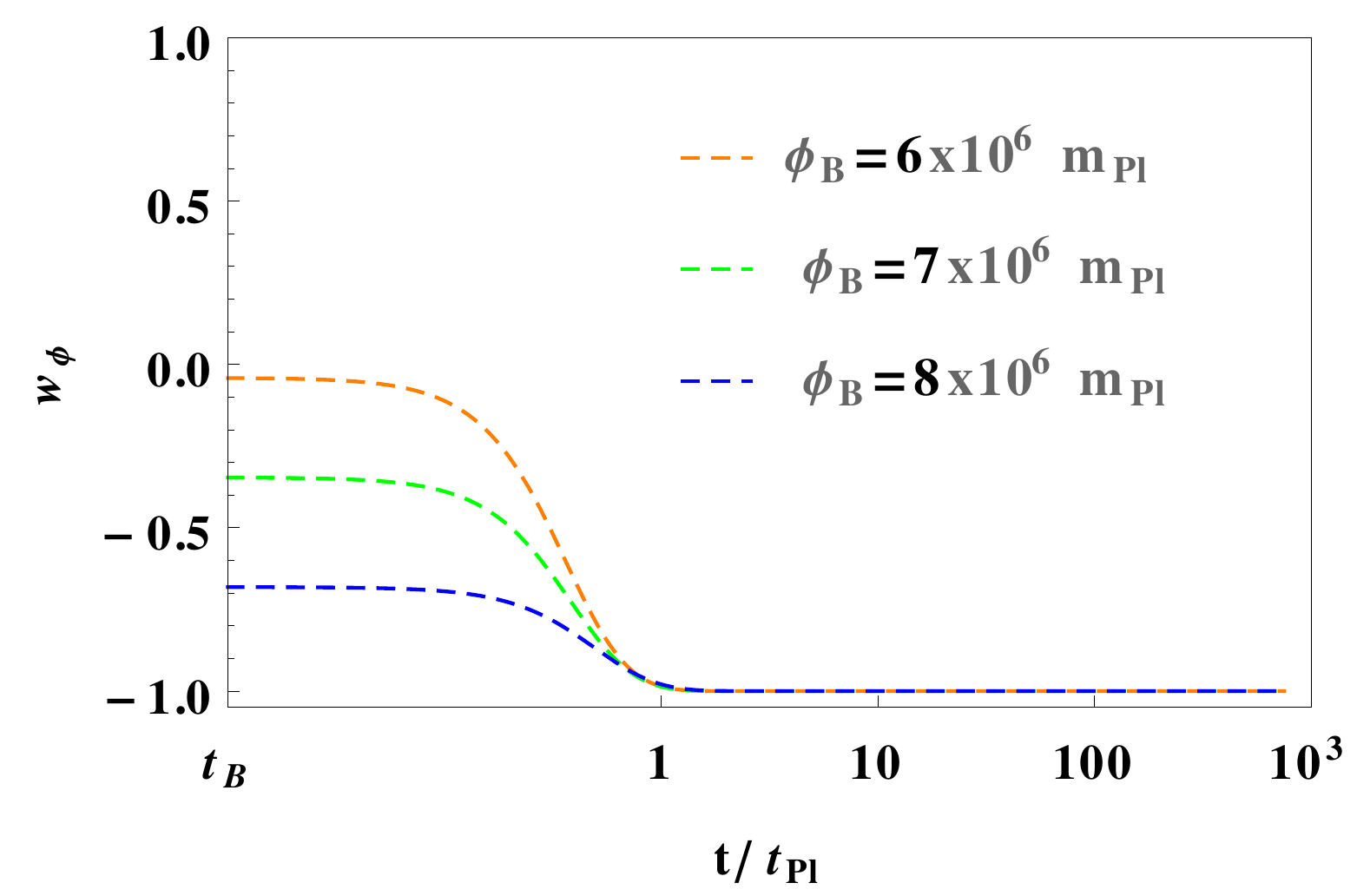}} & 
{\includegraphics[width=0.46\textwidth]{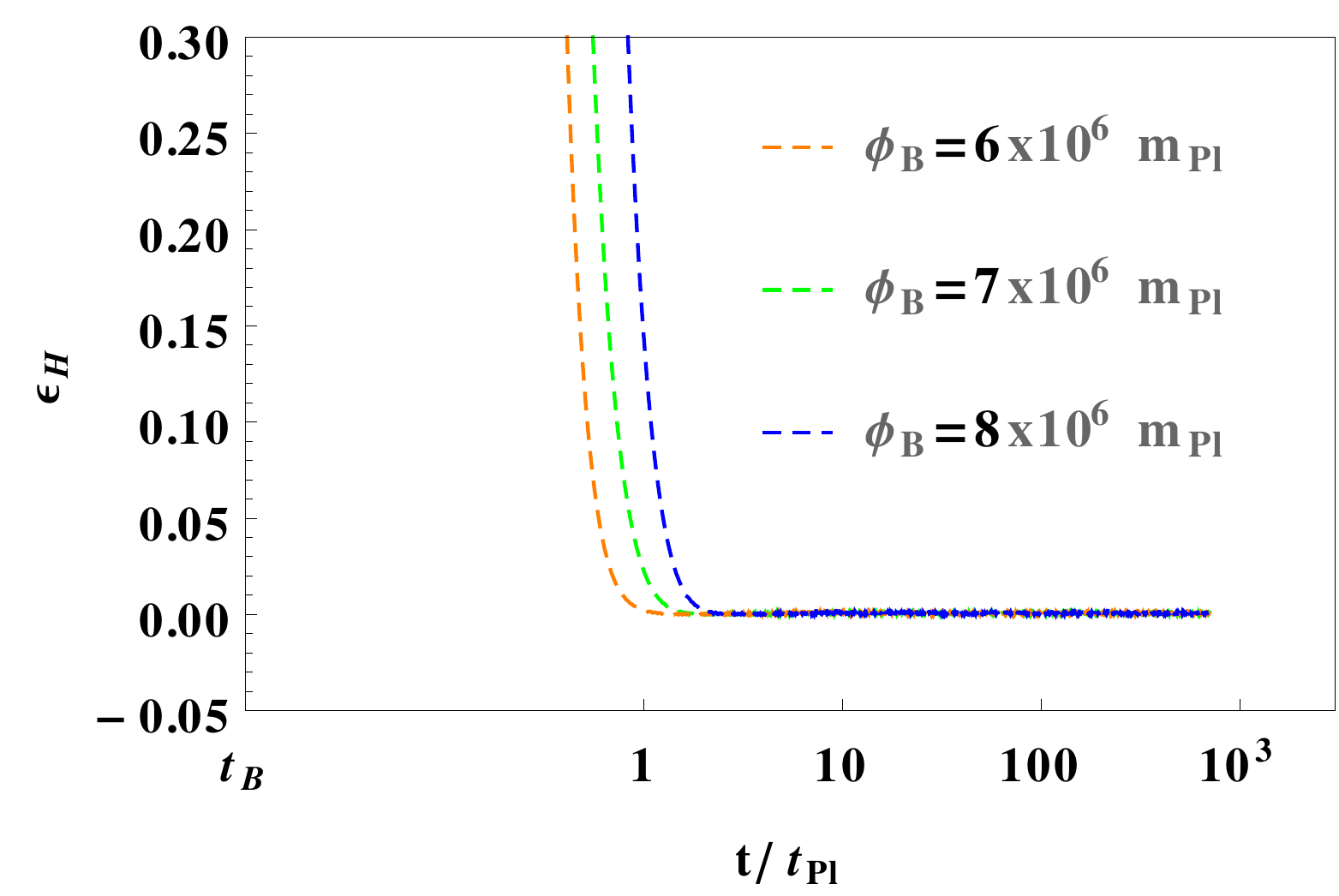}} \\\vspace{6pt}
{(\textbf{e})} & {(\textbf{f})}
\end{tabular}
\end{center}\vspace{-10pt}

\caption{ The numerical results for the potential (\ref{eq:pot}) with $n=5/3$ and $\dot{\phi}_B>0$. This figure exhibits the numerical evolution of $a(t)$, $w(\phi)$, and $\epsilon_H$ in case of the ((\textbf{a}), (\textbf{b}), (\textbf{c})) kinetic energy-dominated (KED)  and ((\textbf{d}), (\textbf{e}), (\textbf{f})) potential energy-dominated (PED)  initial values at the quantum bounce. The analytical solution of the scale factor (Equation~(\ref{eq:a})) is also displayed in order to compare it with the numerical solution shown in the (\textbf{a}) and (\textbf{d}). In the KED case (\textbf{a}), the evolution of $a(t)$ shows universal feature while it disappears in the PED case (\textbf{d}). We chose $m= 10^{-5}m_{pl}$ and $m_{pl}=1$. One can also obtain a similar figure with $\dot{\phi}_B<0$.}%1. Please check if (a-f) is acceptable.  2. Please check ``Top-left panel'' into ``a'' or others.

\label{fig:5/3a}
\end{figure}

Let us discuss  the $\dot{\phi}_B > 0$ case with the KED and the PED initial values at the bounce. The numerical evolution of $a(t)$, $w(\phi)$, and  $\epsilon_H $ are shown in 
Figure~\ref{fig:5/3a} for the same set of initial conditions of $\phi_B$. In the case of the KED initial values ((a)-(c)), %if it should be changed into (a-c) 
the evolution of the scale factor $a(t)$ is independent for a large range of initial values, and shows a universal feature that is consistent with the analytical solution (\ref{eq:a}). During the slow-roll regime, $a(t)$ increases exponentially. By looking at the (b) panel of Figure~\ref{fig:5/3a}, one~can see that the evolution of the Universe is divided into three regimes: bouncing, transition, and slow-roll. During the bouncing regime, $w(\phi) \simeq +1$. In the transition phase ($t/t_{pl}\simeq 10^4$), $w(\phi)$ changes from $+1$ to $-1$ ($t/t_{pl}\simeq 10^5$). In the slow-roll regime, $w(\phi)$ stays close to $-1$ until the end of the slow-roll inflation. 

In addition, we obtain $N_{inf}$ in the slow-roll regime for different values of $\phi_B$ within the range $(0, \phi_{max})$, and the relevant quantities are presented in Table \ref{tab:5/3}. To be consistent with the present observation, one should get at least 60 $e$-folds that restricts the range of $\phi_B$, as given below:
\begin{eqnarray}
\phi_B \in (0.75 m_{pl}, \phi_{max}) \qquad \text{for}~~~ \dot{\phi}_B>0,
\end{eqnarray}
where
\begin{eqnarray}
\phi_{max} = \left( \frac{8 \rho_c^3}{m^7}  \right)^{1/5} \equiv 8.87 \times 10^6.
\label{eq:phim5/3}
\end{eqnarray}

From Table \ref{tab:5/3}, one can conclude that $N_{inf}$ grows as $\phi_B$ increases, which shows that a large value of $\phi_B$ provides more $N_{inf}$ in the slow-roll regime. {Similar}%was "The alike". Please confirm
~consequences for the power-law potential with $n<2$ were discussed in~\cite{alam2017}. 

Next, we choose the PED initial values at the bounce, and the results are shown in the (d)-(f) panels of Figure~\ref{fig:5/3a}. In this case, the universality of the expansion factor $a(t)$ disappears, and the bouncing region no longer exists, although the slow-roll regime can be achieved. As mentioned above, one can obtain more $N_{inf}$ for the large value of $\phi_B$.

%%%%%%%%%
%%%%%%%%%%%

As we discussed, the scalar field $\phi$ should be positive in order to obtain the real potential. As a result, the phase space trajectories will cover only a half circle. Hence, to display the phase portrait in a whole circle, we consider the dimensionless variables, as given below.
\begin{eqnarray}
\xi=\pm \sqrt{\frac{V(\phi)}{\rho_c}}, \qquad y=\pm \frac{\dot{\phi}}{\sqrt{2 \rho_c}}, \qquad \tau=m t.
\label{eq:dimquant}
\end{eqnarray}

Therefore, an autonomous system of the background equations is given by
\begin{align}
\frac{d \xi}{d\tau}&= \pm \sqrt{2V_0} \frac{n}{2m} \left( \frac{\rho_c \xi^2}{V_0} \right)^{\frac{n-2}{2n}} y \nonumber,\\
 \frac{d y}{d\tau}&= -3Hy/m -  \frac{V_0 n}{m\sqrt{2\rho_c}} \left( \frac{\rho_c \xi^2}{V_0} \right)^{\frac{n-1}{n}},
\label{eq:auto}
\end{align}
where $V(\phi)=V_{0} \phi^n$ and $V_0=1/2~ m^{4-n}$. At the bounce, we have $\rho=\rho_c$, which gives the equation  of a~circle:
\begin{eqnarray}
\xi_B^2 + y_B^2 =1.
\label{eq:circle}
\end{eqnarray}

Hence, for the given dimensionless quantities, one can get the equation of a circle at the quantum bounce which is independent of $n$.

Let us examine the phase portrait analysis for the potential (\ref{eq:pot}) with $n=5/3$. Table \ref{tab:5/3} % is~displayed
{displays $N_{inf}$ for cases of}%Please confirm that your intended meaning is retained.
~$\dot{\phi}_B>0$ and $\dot{\phi}_B<0$. By looking at the table, we conclude that the physical viable initial values are $\phi_B \geq 0.75 m_{pl}$ for $\dot{\phi}_B>0$ and $\phi_B \geq 5.27 m_{pl}$ for $\dot{\phi}_B<0$, which are consistent with the observations. Thus, in the entire region of initial values, these are the only KED initial conditions that can generate the desired slow-roll inflation. However, few initial values do not give the desired slow-roll phase, as shown in Table \ref{tab:5/3}. The whole range of the PED initial values at the quantum bounce can produce the desired slow-roll inflation, and a large $N_{inf}$ can be found.

%%%%%%%%%
\begin{table}
\caption{Table for $N_{inf}$ in case of $\dot{\phi}_B>0$ and $\dot{\phi}_B<0$.}
\centering
\begin{tabular}{lcccl}
\hline\\
$\dot{\phi}_B$ & $\phi_B$  & Inflation & $t/t_{pl}$  & $N_{inf}$ \\\\
\hline
$>0$ & 0 & starts & $6.5216 \times 10^4$ &  40.32 \\
 & & ends & $1.1035 \times 10^7$ &   \\\\
& 0.75 & starts & $5.1754 \times 10^4$ &  60.01 \\
 & & ends & $1.1167 \times 10^7$ &  \\\\ 
 & 0.80 & starts & $5.1070 \times 10^4$ &  65.73 \\
 & & ends & $1.7140 \times 10^7$ &   \\\\ 
 & $8 \times 10^6$ & starts & $0.577$ &  709.81 \\
 & & ends & $1.0473 \times 10^3$ &   \\\\ 
$<0$ & 4.5 & starts & $6.4189 \times 10^4$ &  37.87 \\
 & & ends & $1.0699 \times 10^7$ &   \\\\ 
& 5.27 & starts & $4.9773 \times 10^4$ &  60.32 \\
 & & ends & $1.1113 \times 10^7$ &   \\\\  
 & 6.0 & starts & $4.1424 \times 10^4$ &  78.21 \\
 & & ends & $1.1040 \times 10^7$ &   \\\\
\hline
\end{tabular}
\label{tab:5/3}
\end{table}
%%%%%

%%%%%

\begin{figure}
\begin{center}
\begin{tabular}{c}
{\includegraphics[width=0.46\textwidth]{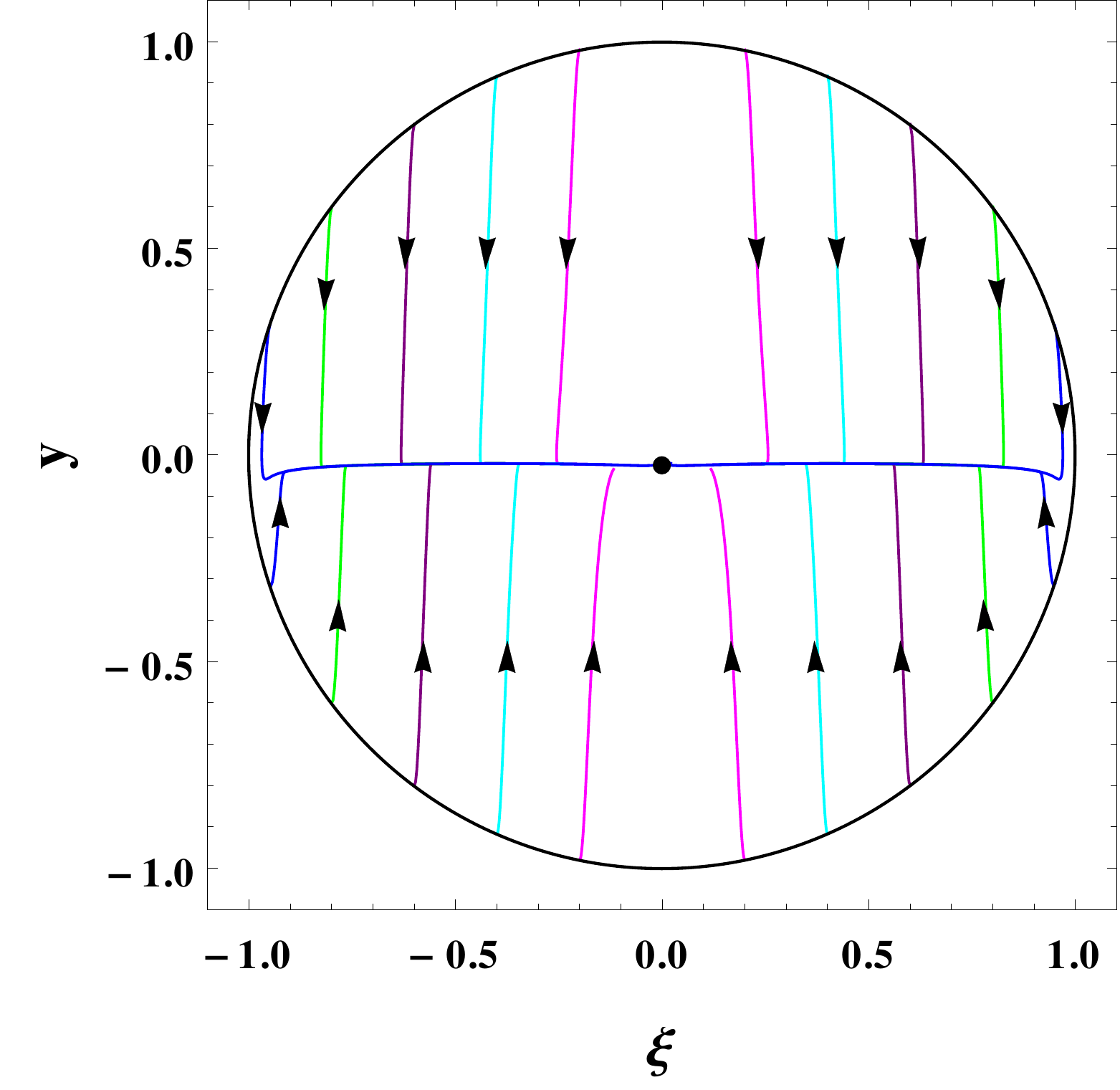}} 
\end{tabular}
\end{center}\vspace{-12pt}
\caption{The phase portrait for the potential (\ref{eq:pot}) with $n=5/3$. The trajectories with arrows exhibit the time evolution from the bouncing boundary to the origin. We chose $m=0.2 m_{pl}$ for better depiction.}
\label{fig:portrait}
\end{figure}

Figure~\ref{fig:portrait} exhibits the numerical evolution of the phase portrait for $\dot{\phi}_B>0$ and~$\dot{\phi}_B<0$,~together with the KED and the PED initial values that are started from the bounce. Regions near the circle correspond to the maximum energy density, while regions close to the origin represent minimum energy density in the $\xi-y$ plane. All the trajectories onset from the bounce, and meet at the inflationary separatrix before moving towards the origin~\cite{psingh2006}. 

Finally, we compare our results with the quadratic and Starobinsky models~\cite{Tao2017,Bonga2016}. Our analyses are in agreement with the quadratic potential. However, the quadratic potential has been practically disfavored by the Planck data~\cite{Planck2015}. In the case of $N_{inf}$, both the KED and the PED initial values are compatible with observations for the potential with $n=5/3$, whereas the Starobinsky potential favors only  KED initial values and not  the PED ones at the bounce~\cite{Tao2017,Bonga2016}.

%%%%%%
%%%%%%%%%

%%%%%%%%
\section{Conclusions}
\label{sec:conc}
We examined the pre-inflationary Universe of the potential (\ref{eq:pot}) with $n=5/3$ in the framework of LQC. We restricted ourselves to the positive values of inflaton field for the real potential. In the case of the KED initial values, the Universe is always divided into three regimes: {\em bouncing, transition,} and {\em slow-roll inflation}. During the bouncing regime, the background evolution is independent for a wide range of initial values and also the potential. Specially, the numerical evolution of the expansion factor  represented the universal characteristic and compared by the analytical solution (\ref{eq:a}), as~shown in Figure~\ref{fig:5/3a}. In this phase, $w(\phi) \simeq +1$, whereas in transition regime, it decreases drastically from $w(\phi) \simeq +1$ to $-1$. The period of the transition regime is very small in comparison with the other two regimes. Later, the Universe goes into an accelerating phase, where $\epsilon_H$ is still large, but it rapidly goes to zero where the slow-roll inflation begins (see Figure~\ref{fig:5/3a}). Thereafter, we found $N_{inf}$ during the slow-roll inflation, as displayed in Table \ref{tab:5/3}. For the PED initial values, the universal characteristic  of the scale factor $a(t)$ disappeared, and the bouncing phase does not occur, but the slow-roll inflation can still be obtained for a long period, and correspondingly a maximum number of $e$-folds are found which are presented in the (d)-(f) panels of Figure~\ref{fig:5/3a} and Table \ref{tab:5/3}. 

The potential under consideration generates the desired slow-roll inflation in the case of both KED and PED initial conditions, and produces enough $e$-folds that are compatible with observations, whereas Starobinsky and $\alpha-$attractor (dependence on the range of $\alpha$) models provide the slow-roll inflation only for KED initial conditions and not for PED ones~\cite{Bonga2016,Tao2017,alam2018}.

We also displayed the phase portrait for the model under consideration, and the phase space trajectories are designated in Figure~\ref{fig:portrait}. In the phase portrait, the trajectories start from the bounce and go towards the origin asymptotically for a generic set of initial conditions.

%%%%%%%%%%%%%%%%%%%%%%%%%%%%%%%%%%%%%%%%%%
\vspace{6pt}

\newpage
%=====================================
% References, variant A: internal bibliography
%=====================================
%\reftitle{References}

\end{document}